# A Framework for Constraint-Based Deployment and Autonomic Management of Distributed Applications (Extended Abstract)

Alan Dearle, Graham N.C. Kirby and Andrew J. McCarthy

*School of Computer Science, University of St Andrews, St Andrews, Fife KY16 9SS, Scotland*
{al, graham, ajm}@dcs.st-and.ac.uk

## Abstract

*We propose a framework for the deployment and subsequent autonomic management of component-based distributed applications. An initial deployment goal is specified using a declarative constraint language, expressing constraints over aspects such as component-host mappings and component interconnection topology. A constraint solver is used to find a configuration that satisfies the goal, and the configuration is deployed automatically. The deployed application is instrumented to allow subsequent autonomic management. If, during execution, the manager detects that the original goal is no longer being met, the satisfy/deploy process can be repeated automatically in order to generate a revised deployment that does meet the goal.*

We believe that the initial deployment of an application and its subsequent evolution in the face of host failures and other perturbations are separate but closely related problems. Both are too complex in large applications to be handled by a human operator. We propose that both should be controlled automatically, driven by a high-level configuration goal specified by the administrator at the outset. We thus address specifically the first and third of Kephart & Chess' issues [1]: self-configuration and self-healing.

Our general approach is shown below. The application administrator specifies a deployment **goal** in terms of resources available and constraints over their deployment. We propose a new domain-specific constraint language called **Deladas** (DEclarative LAnguage for Describing Autonomic Systems) for this purpose. The resources include software components and physical hosts on which these components may be installed and executed. Constraints operate over aspects such as the mapping of components to hosts and the interconnection topology between components.

The autonomic cycle is controlled by an engine, which we call the Autonomic Deployment and Management Engine (ADME), that attempts to **satisfy** a goal, specified by the administrator in the constraint language. The engine includes a parser and constraint solver. The result of the attempted goal satisfaction is a set of zero or more solutions. Each solution is in the form of a **configuration**, expressed as a Deployment Description Document (DDD), which describes a particular mapping of components to hosts and interconnection topology that satisfies the constraints.

If a configuration can be found, it is **enacted** by the engine to produce a running deployment of the application. From a DDD, the ADME generates a collection of scripts which perform installation, instantiation and wiring of the components. Once the scripts have executed on the appropriate hosts, the application is fully deployed in its initial configuration [2].

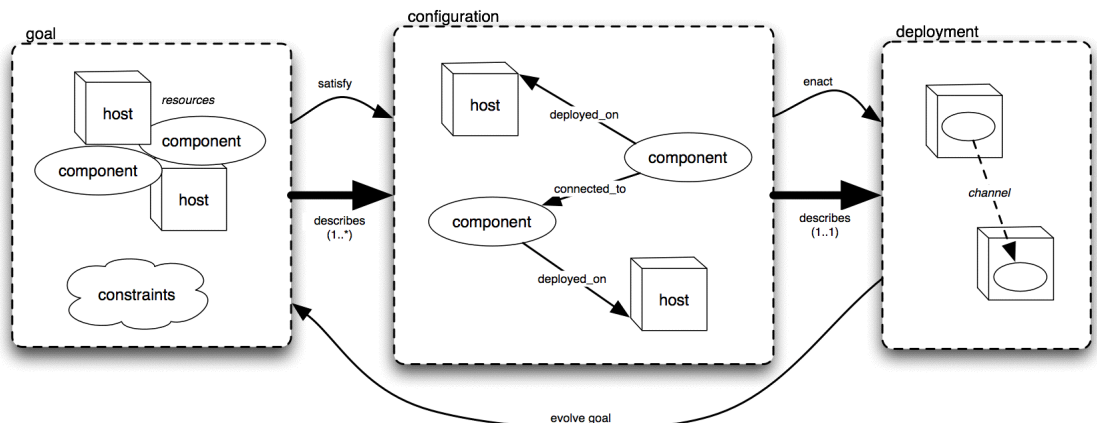

The autonomic aspect of this approach is that the deployed application is instrumented with probes to monitor its execution. Events generated by the probes are sent to the ADME, which may decide that the deployment no longer satisfies the original goal, for example if a component or host fails. In this case the ADME evolves the goal to take account of changed resource availability—for example, removing failed hosts and perhaps adding new hosts that may now be available—and initiates the **satisfy/enact** cycle again. This attempts to find a new solution of the constraints that combines existing and new components, and to enact this in an efficient manner. Assuming that such a new configuration can be found and deployed, the system has reacted automatically and appropriately to a change in the application's environment. The cycle may continue indefinitely.

The nature of the probes required to monitor the application depends on the constraints specified in the goal. At the simplest level the constraints operate over just the component/host topology, and for this, simple probes are sufficient. Where more complex probes are required, this can be deduced by ADME from the specified constraints. For example, constraints can operate over the latency or bandwidth of a channel, the degree of replication of a component, or the mean availability of a host. Each of these dynamic aspects requires a specialised probe. Deladas may be extended to incorporate new constraint types and probes.

This style of autonomic application evolution can be achieved without human intervention. The framework also accommodates the need for more wide-ranging evolution. For example, in addition to changes in the application's environment, changes may occur in the supported enterprise, requiring manual revision of the deployment goal, including the constraints.

To illustrate the use of Deladas, we use an example drawn from the peer-to-peer domain, in which *clients* connect to *routers*. In the example code, the *constraintset* contains five constraint clauses. These clauses operate over two types of component named *Router* and *Client*. It is not necessary to specify the concrete types of these components but it is possible to infer that, in order to satisfy the constraints, the component *Router* must have ports named *rin*, *rout*, *cin* and *cout*. The constraints are written in first-order logic and specify (in sequence) that:

- hosts run an instance of a router and/or a client
- each client connects to at least one router via *out* and *in* ports and *cin* and *cout* ports respectively
- there are at most two clients for every router
- every router is connected to at least one other router via their *rin* and *rout* ports
- routers are strongly connected

```
constraintset randc = constraintset {
  // 1 router or client per host
  forall host h in deployment (
    card(instancesof Router in h) = 1 or
    card(instancesof Client in h) = 1 )
  // every client connects to at
  // least 1 router
  forall Client c in deployment (
    exists Router r in deployment (
      c.out connectsto r.cin
      c.in connectsto r.cout ))
  // every router connects to at
  // most 2 clients
  forall Router r in deployment (
    card(Client c connectedto r) <= 2 )
  // every router connects to at
  // least 1 other router
  forall Router r1 in deployment (
    exists Router r2 in deployment (
      r1.rout connectsto r2.rin
      r1.rin connectsto r2.rout
      r1 != r2 ))
  // routers are reachable from each other
  forall Router r1,r2 in deployment (
    reachable(r1, r2))
}
```

In conclusion, this abstract has outlined a framework to support the initial deployment and subsequent autonomic evolution of distributed applications in the face of perturbations such as host and link failure, temporary bandwidth problems, etc. The knowledge required for autonomic management is specified in the form of a set of available hardware and software resources and a set of constraints over their deployment. We postulate that it is feasible to implement an autonomic manager that will automatically evolve the deployed application to maintain the constraints while it is in operation. We are currently working on an implementation to enable us to test this assertion. A full version of this paper is available [3].

This work is supported by EPSRC Grants GR/M78403, GR/R51872, GR/S44501 and by EC Framework V IST-2001-32360.